# Simulating the Influence of Collaborative Networks on the Structure of Networks of Organizations, Employment Structure, and Organization Value

Willy Picard,

Poznań University of Economics,
Department of Information Technology,
al. Niepodległości 10, 61-875 Poznań,
picard@kti.ue.poznan.pl

**Abstract.** From the perspective of reindustrialization, it is important to understand the evolution of the structure of the network of organizations employment structure, and organization value. Understanding the potential influence of collaborative networks (CNs) on these aspects may lead to the development of appropriate economic policies. In this paper, we propose a theoretical approach to analysis this potential influence, based on a model of dynamic networked ecosystem of organizations encompassing collaboration relations among organization, employment mobility, and organization value. A large number of simulations has been performed to identify factors influencing the structure of the network of organizations employment structure, and organization value. The main findings are that 1) the higher the number of members of CNs, the better the clustering and the shorter the average path length among organizations; 2) the constitution of CNs does not affect neither the structure of the network of organizations, nor the employment structure and the organization value.

## 1 Introduction

Industrialization has given birth to an economic world in which manufacturing of goods and services is the base of the society. In many industrialized societies, major economic changes, such as globalization and specialization, have led to a shift from an industrial economy to a more service-oriented one. This evolution is referred to as *deindustrialization*. A commonly accepted definition of deindustrialization is "the decline in importance of manufacturing industry in the economy of a nation or area" [1]. Two processes are at work during deindustrialization [2]: on the one hand, specialization of organizations on their core competences increases their productivity. Next, this increase of productivity leads to a reduction of the quantity of human resources needed to create a given value. Therefore, less people are needed in the industry. On the second hand, the wages are superior in the service sector than in the industry. As a consequence, a reallocation of workers from the industry to the service sector may be observed. These two processes combined together lead to a decrease in the number of employees working in the industry, associated with a reduced number of industrial organizations. In parallel, there is a rise of the number of organizations and employees in the service sector [3].

It has been argued that deindustrialization has dramatic effect on many economic and social aspects of local, regional, national and transnational economies. Besides the rise of unemployment, deindustrialization leads to strong limitations with regards to innovativeness: the lack of industrial partners to design, develop, and prototype new products is a strong brake to innovation [4].

Therefore, in many post-industrial countries, the process of *reindustrialization* is currently under scrutiny. Reindustrialization is the renewed development of the industrial sector. Although the idea of reindustrialization is not new, the rise of the economic importance of the People's Republic of China, especially with regard to the industrial sector, has given rise to a debate on the importance of reindustrialization.

The question of the importance of reindustrialization is often related to employment which is still at the level of 10.12% in 2010 in the Euro area, 8,18% among the G7 members, while at the level of 4.1% in the People's Republic of China, according to the International Monetary Fund [5]. In this context, there is a need for a good understanding of the variables that may influence the reallocation of workers. Pastore [6] has studied the relation between the level of unemployment at the regional level and the reallocation rate in Italy. Worker reallocation in Canada have been studied by Morissette, Lu and Qiu [7], taking into account many variables, such as workers' age, organization activity sector, organization size. A similar study has been published by Liu about China [8]. Martin and Scarpetta [9] have a different approach to the question of employment and reindustrialization: they have addressed the question of the links between regulations for employment protection, workers' reallocation, and productivity. Bartolucci and Devicienti [10] have demonstrated that better workers are found to have a higher probability of moving to better firms. Gianelle [11] has studied in details the structure of workers reallocation in the Veneto industrialized region of the north of Italy. He has shown the importance of hub organizations, i.e., highly connected organizations bridging distinct local clusters of organization, for workers' mobility. In this work, connections among organizations represent reallocations of employees: each reallocation of an employee from an organization $O_1$ to an organization $O_2$ is represented as a link $O_1 \rightarrow O_2$ between these organizations.

None of these studies have considered the potential influence of cooperation among organizations on the dynamics of employment. Specialization, which is one of the two pillars of deindustrialization, is also a key reason for the decision of some organizations to collaborate with other organizations. By collaborating with other organizations, creating a *collaborative network (CN)*, on the one hand, organizations focus on their core competences and benefit from their competitive advantage, while, on the other hand, the group of collaborating organizations, i.e., the CN, is able to produce complex products or provide complex services. Intuitively one may expect a relation between the existence of CNs and reallocation: workers are probably more willing to move to organizations that they have already collaborated with, than to move to unknown ones. However, to our best knowledge, the relation between the existence of CNs and the dynamics of employment has never been studied.

In this paper, we take a theoretical approach to this question. We propose a model of the dynamics of employment in a networked ecosystem of organizations tied by their collaboration in CNs. We further use the model to run simulations to evaluate the influence of CNs, especially their size and their constitution, on the structure of the network of organizations, on employment and on the value of the organizations themselves

In Section 2, background on network structure and chosen metrics, i.e., clustering coefficient and degree distribution, is provided. Next, the proposed model is detailed in Section 3. In Section 4, the results of the simulation experiment are presented. In Section 5, the results are discussed. Finally, Section 6 concludes the paper.

## 2  Background on Networks

The word "network" refers to the notion of a set of interconnected objects, referred to as nodes, that may be material or immaterial. Networks are ubiquitous and have been the object of research in various disciplines, such as computer science, sociology, biology, and medicine to name a few [12-14].
It has been demonstrated that many of these networks share a set of common characteristics: their average shortest path is relatively low, therefore two nodes of the network are connected by a small number of links. Additionally in many networks, the immediate neighbors of a node tend to be connected to each other, i.e., their clustering coefficient is higher than in random networks. Such networks are referred to as "small-world networks" [15,16] and their structure and functioning have been the object of numerous works since Watts and Strogatz's article [15] published in 1998. Networks have also been studied with regard to their dynamics. As an example, network percolation, with its potential application to explain disease epidemics and gossip propagation, has received significant attention [17-19].

Among key characteristics of networks, the *clustering coefficient*, the *scale-free* and the *small-world* properties of some networks have been intensively scrutinized. The clustering coefficient is a measure of the tendency of nodes to connect to other nodes within groups of nodes. Formally, the clustering coefficient $C$ of a graph $G = (V, E)$ has been defined by Watts and Strogatz as follows: "Suppose that a vertex $v$ has $k_v$ neighbours; then at most $k_v(k_v - 1)/2$ edges can exist between them (this occurs when every neighbor of $v$ is connected to every other neighbour of $v$). Let $C_v$ denote the fraction of these allowable edges that actually exist. Define $C$ as the average of $C_v$ over all $v$." The higher the clustering coefficient, the higher the number of connected triplets of nodes. Scale-free networks are networks in which the probability that a randomly selected node has $k$ links, i.e., degree $k$, follows $P(k) \sim k^{-\gamma}$, where $\gamma$ is the degree exponent. In scale-free networks, a limited number of nodes have a large number of links (a large degree), while a large number of nodes have a small number of links (a small degree). Small-world networks are graphs in which the clustering coefficient is high and the average path length is small. In small-world networks, a relatively small number of nodes separate any two of them, even if most nodes are not connected to each other.

Popular models of structure of networks are:
- the Erdős–Rényi model: this model is a random model in which the links between nodes are added in a random manner. Both clustering and average path length of Erdős–Rényi networks are low;
- the Barabàsi-Albert model: in this model, a newly added node is connected to other nodes of the network, such that the probability to connect to a given node is proportional to the degree of this node. Barabàsi-Albert networks are scale-free and their clustering coefficient is higher than in Erdős–Rényi networks.

More information about social networks analysis may be found in [20].

## 3  A Model of a Dynamic Networked Ecosystem of Organizations

### 3.1  Assumptions of the Proposed Model

The proposed model is based on the following set of five assumptions.

*Assumption 1 – **Synergy**: Collaboration among organizations increases their value.*

The synergy assumption is based on the idea that having organizations collaborating towards the achievement of a common goal by sharing their competences brings an added value to each of the collaborating organizations. So some extends, this assumption lies at the bottom of the concept of collaborative networks.

*Assumption 2 – **Erosion**: The part of the value of an organization originated by former collaboration with other organizations fades out when this collaboration ends.*

The erosion assumption takes on the idea that when organizations are not collaborating anymore, the synergy value created during their collaboration time, sustains in time. However, this synergy value vanishes over time. Therefore, the synergy value created by collaboration is fading away as the employees are not collaborating any longer.

*Assumption 3 – **Specialization advantage**: The value created by an employee is maximal when (s)he has the same profile as the organization that employs her/him.*

The specialization advantage assumption is based on the idea that when an employee is working in an organization with a different profile that his/her own, the adaptation of the employee's professional culture and how-to to the culture and how-to of the organization comes with a cost. For instance, an IT professional is less efficient when working at a pharmaceutical organization than at a software organization.

*Assumption 4 – **Local preference**: an employee quitting a job preferably moves to an organization that has a collaboration history with the left organization.*

The local preference assumption is related with the works on job mobility by Holzer [21] and Bewley [22], who have identified that 53% (resp. 60%) of the employers are seeking future employees on the social networks of their employees.

*Assumption 5 – **Profile preference**: an employee quitting a job preferably moves to an organization whose profile is that same as her/his own.*

The profile preference assumption is related to the idea that the specialization advantage leads to a higher efficiency of the employee, potentially related to higher wages. For instance, an IT professional will rather move to a software organization sharing his/her culture, instead of a pharmaceutical organization.

### 3.2 A Model of a Networked Ecosystem of Organizations

In the proposed model, an ecosystem of organizations is modeled as a triplet: a *network of organizations*, an *employment structure*, and a *set of collaborative networks*. In the network of organizations $N$, the nodes represent the organizations, while the links represent the collaboration history of organizations. Let $O$ denote the set of organizations of the ecosystem, with $o_i$ being the $i$-th organization. The total number of organizations in the ecosystem is assumed to be fixed.

Let $l_{ij} = l_{ji} \in L$ denote the link between organizations $o_i$ and $o_j$ representing collaboration between organizations $o_i$ and $o_j$ collaborate with each other. The network of organization is undirected as the collaboration relation is symmetric, i.e., if the organization $o_i$ collaborates with the organization $o_j$, then the organization $o_j$ collaborates with $o_i$. The network of organizations is then $N = \{O, L\}$.

The employment structure $E$ of a networked ecosystem of organizations is modeled as a matrix of dimensions $(|O| + 1) \times |P|$, where $|O|$ is the number of organizations in the ecosystem, and $|P|$ is the number of profiles of employees in the ecosystem. $P = \{p_1, \dots, p_{|P|}\}$. Examples of profile are "IT professional" and "Accountant". The elements of the employment structure $E$ are positive integers, such that,

$$e_{ij} = \begin{cases} \text{the number of unemployed people of profile } p_j, \text{if } i = 0, \\ \text{the number of employees of the organization } o_i \text{ with profile } p_j, \text{otherwise.} \end{cases}$$

The total number of employees in the ecosystem is assumed to be fixed.

The profile of an organization is defined as the predominant profile of its employees, i.e., $p_{o_i} = p_j \Leftrightarrow e_{ij} = \max(e_{ik})$, for $k \in \{1, \dots, |P|\}$.

Let $CN$ denote the set of collaborative networks. Each collaborative network is a triplet $cn_i = \{\theta_i, CO_i, d_i\}$, where $\theta_i$ is the point in time in which the collaborative network has been created, $CO_i$ is the set of organizations collaborating within the collaborative network, and $d_i$ is the duration of the collaborative network. Therefore, the collaborative network is dissolved at $t = \theta_i + d_i$.

### 3.3 Organization value

Finally, let $v_i$ denote the value of the organization $o_i$. The value of an organization should encompass both the value created by its employees and the value created by synergy with other organizations. Let $v_i^e$ denote the value created by the employees of the organization $o_i$, and $v_i^s$ the value created by the synergy with other organizations.

The value created by employees supports the specialization advantage assumption. In our model, the value brought by an employee to the value of an organization with the same profile is normalized to 1, while the value brought by an employee to the value of an organization with a different profile equals $0 < \rho < 1$. The total value of an organization created by its employees is then $v_i^e = \rho \sum_{\substack{i \neq 1 \\ i \neq p_o}} e_{ij} + \sum_{i=p_o} e_{ij}$.

The value created by the synergy with other organizations has to support both the synergy and the erosion assumptions. The synergy assumption implies that each new collaborative network within which an organization is collaborating increases the value of the organization. The erosion assumption assumes that when an organization is not collaborating with any other organization, its synergy value is fading out. The synergy value created by an organization $o_i$ collaborating with the organization $o_j$ equals to $v_{ij}^s = 1 + \log(|CN_i|)$, where $CN_i$ is the subset of collaborative networks in which the organization $o_i$ participates. Therefore the synergy value of an organization collaborating with another organization in one collaborative network equals $1 + \log(1) = 1$, the synergy value in 5 collaborative networks equals $1 + \log(5) \sim 2{,}6$.

When an organization is not participating in any collaborative network, its synergy value is decreasing in time. Let $t_{\omega_{ij}}$ denote the last time when the organization has participated to a collaborative network with organization $o_j$, with potentially $t_{\omega_{ij}} = 0$ for organizations that have never participated to any collaborative network. In our model, the synergy value is defined as follows:

$$v_{ij}^s(t) = 1 - \frac{1}{1+e^{f_s \cdot \left(\frac{f_d}{2}+t_{\omega_{ij}}-t\right)}}, \quad (1)$$

where $f_s$ is the fading slope, and $f_d$ is the fading duration. Additionally, a threshold $f_t$ is defined to filter out old, insignificant relations among organization. Therefore, if $v_{ij}^s(t) < f_t$, then the link $l_{ij}$ is removed from the set of links $L$. The function defined by the Eq. 1 and threshold filtering are illustrated in Fig. 1.

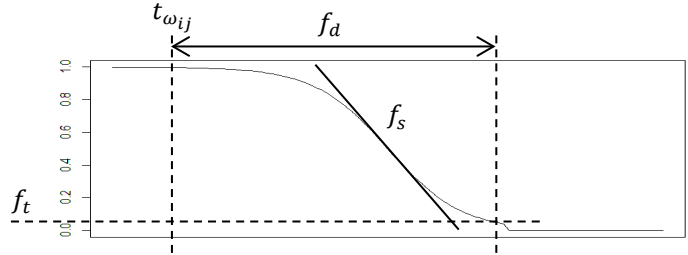

**Fig. 1.** The fading function (Eq. 1) of the synergy value for organizations that are not participating in collaborative networks.

Next, the total value created by the synergy with other organizations may be calculated on the bases of the values $v_{ij}^s$. In out model, we distinguished collaboration between same-profile organizations from collaboration between organizations having different profiles. In terms of synergy, collaboration between organizations having different profiles is more valued than collaboration between same-profile organizations.

Therefore, with $0 < \theta < 1$, $v_i^s = \theta \sum_{\substack{j \neq i \\ p_{o_i}=p_{o_j}}} v_{ij}^s + \sum_{\substack{j \neq i \\ p_{o_i} \neq p_{o_j}}} v_{ij}^s$.

Finally, the total value of an organization is defined as $v_i = v_i^e + \alpha v_i^s$, where $\alpha$ is the weight of the synergy value with regard to the value created by the employees.

## 4   Dynamic Networked Ecosystem of Organizations

### 4.1   Employees Mobility in a Networked Ecosystem of Organizations

The dynamics of employees is modeled with three basic operations: *hire* (an unemployed person is hired by an organization), *fire* (an employee is fired by his/her organization), and *quit* (an employee is moving to another organization).

In the proposed model, we do not consider employees which are removed from the labor market, for various reasons, including retirement, death, and accidents leading to severe disability. The hiring rate is proportionally dependent on the ratio of the value of an organization to the number of employees. The higher this ratio is, the greater the productivity of the organization that may want to hire more employees. Similarly, the firing rate is proportionally dependent on the ratio of the number of employees to the value of an organization. The higher this ratio is, the lesser the productivity of the organization that may want to fire more employees. The quitting rate is different than the hiring and firing rate, as it is an employee's decision, instead of an employer's decision. In our model, the quitting rate is proportional to the ratio between the mean value of the neighbors of the organization in which the employee works to the value of this organization. Therefore, if the mean value of the neighboring organization is higher than the value of the organization in which the employee is working, i.e., "surrounding" organizations are doing better than the employee's organization, than the employee is more willing to quit.

In the proposed model, the choice of the organization in which the quitting employee will work has to encompass both the local and the profile preferences. Therefore, a quitting employee should prefer local organizations, i.e., organizations in the neighborhood of his/her current workplace, over other, distant organizations. Here, neighbor organizations do not have to be geographically close, but they have to be tied with regards to collaboration. Additionally, a quitting employee should prefer organizations with which (s)he shares the same profile. These preferences are captured in our model by $\pi_l$ and $\pi_p$ being the probability for a quitting employee to choose a local organization and the probability to choose a same-profile organization.

### 4.2   Dynamics of Collaborative Organizations

In the proposed model, collaborative organizations are created according to the following algorithm. At each discrete time moment $t$, each organization has the possibility to create a collaborative organization with a probability $\pi_{CN}$. The duration of each new collaborative organization is chosen randomly, according to the uniform distribution between $d_{\min}$ and $d_{\max}$. The number of collaborators is also chosen randomly, according to the uniform distribution between $CN_{\min}$ and $CN_{\max}$. The set of potential members of the collaborative network consists of the neighbors of order 2 of the organization, that is the immediate neighbors of the organization as well as their neighbors. Additionally, randomly picked organizations are added to the set of potential collaborators, with a probability $\pi_{\text{random}}$. Finally, the members of the collaborative networks are picked from the set of potential members formerly identified. The given percentage $\pi_{\text{same}}$ of the members of the collaborative network shares the same profile with the creating organization, while $1 - \pi_{\text{same}}$ of the members of the collaborative network has a different profile than the creating organization. A collaborative network created at time $\theta_i$ with a duration $d_i$ is active, i.e., participates actively to the synergy value for all $\theta_i \leq t \leq \theta_i + d_i$. After the time $\theta_i + d_i$, the collaborative network is dissolved and does not participate directly to the synergy value of the organization.

## 5    Simulations

The proposed model has been implemented in the R language and environment [23]. The following independent variables are manipulated in our experiment:
-   the number of members of collaborative networks $d_{\text{min}}$ and $d_{\text{max}}$,
-   the probability of the presence of randomly picked members of collaborative networks $\pi_{\text{random}}$,
-   the percentage $\pi_{\text{same}}$ of the members of the collaborative network sharing the same profile with the creating organization.

The dependent variables to be measured are:
-   the clustering coefficient of the network of organizations,
-   the average path length of the network of organizations,
-   the employee distribution,
-   the organization value distribution.

Two starting configuration of the ecosystem has been tested: in 50% of our simulations, the starting ecosystem was an Erdős–Rényi network; in the remaining 50% of our simulations, the starting ecosystem was a Barabàsi-Albert network. In both cases, the starting ecosystem contains 100 organizations and 197 links.

Three profiles have been used during our simulations, referred to as "red", "blue", and "green". The initial population has been created as follows: for each company, the number of employees for each profile is following the normal distribution with a zero mean and a standard deviation equals 70. Such a distribution creates a high number of micro and small enterprises. The type of each organization is defined at starting time. The number of unemployed people is set as 3% of the number of employees at starting time and is equally distributed among all the profiles.

For each configuration of the model, 25 different ecosystems have been generated. For each generated ecosystems, 500 iterations of the system have been performed before measuring the values of the dependent variables. The measured dependent variables have been further aggregated by processing their geographic mean.

We have found (cf. Table 1) that larger CNs in terms of number of their members (increases of $d_{\text{min}}$ and/or $d_{\text{max}}$) leads to an increase of the clustering coefficient and a decrease of the average path length. As a consequence, larger CNs lead to a small-world of organizations, in which, on one hand, organizations are working in cluster, i.e., the collaborators of an organization are usually collaborating with each other, on the other hand, organizations are relatively close each other, i.e., an organization can reach any organization by a small number of collaborators of collaborators.

Large CNs leads also to the emergence of organizations with a large number of employees. It can be explained by the local preference assumption of our model, which encourages the mobility towards neighboring companies, and therefore, organizations that have a large number of collaborators are more willing to be the destination of quitting employees than organizations with few collaborators. The rise of the organization value is not surprising as an important part of the organization value depends directly on the number of its collaborators.

Modifications of $\pi_{random}$ and $\pi_{same}$ have no influence on the structure of the network of organizations in terms of clustering coefficient. The average path is influenced by $\pi_{random}$, as adding links to non-neighboring organizations creates "short-cuts" in the network of organizations. The number of employees per organization and the organization value are influenced by neither $\pi_{random}$ nor $\pi_{same}$.

**Table 1.** Influence of the size and composition of CNs on the structure of the network of organizations, on employment and on the value of the organizations themselves.

| Dependent variables | Clustering coefficient | Average path length | Employees per organization | Organization value |
|---|---|---|---|---|
| $d_{max}$ | ↗ | ↘ | ↗ | ↗ |
| $d_{min}$ | ↗ | ↘ | → | ↗ |
| $\pi_{random}$ | → | ↘ | → | → |
| $\pi_{same}$ | → | → | → | → |

## 6  Conclusions

A major contribution presented in this paper is a theoretical model of ecosystems of organizations, encompassing employee mobility and collaborative networks. Our simulations based on this model provide some original insights concerning the influence of collaborative networks on the structure of the network of organizations, on employment and on the value of the organizations themselves. We have found that the size of the CNs operating on the business ecosystems has an importance influence on the structure of network of organizations: larger CNs lead to a small-world network of organizations. The constitution of CNs does not affect neither the structure of the network of organizations, nor the employment structure and the organization value.

Among future works, there is an important need to confront the results of our simulations with data from real-world cases. However, a major obstacle to this confrontation is the lack of dataset concerning collaborative networks and their employment structure. Another interesting research area is the potential application of the proposed model to simulate the mobility of employees and understand the potential influence of collaborative networks on this mobility.